\documentclass[aps,nofootinbib,showpacs,amsmath,amssymb, preprint]{revtex4}
\usepackage{amssymb}
\usepackage{amsmath}
\newcommand{\be}{\begin{equation}}
\newcommand{\ee}{\end{equation}}
\newcommand{\bea}{\begin{eqnarray}}
\newcommand{\eea}{\end{eqnarray}}
\begin{document}

~\vskip0.2cm

\title{Asymptotic power series of field correlators
\footnote{Presented at the International Conference
 "Selected Topics in Mathematical and Particle Physics" organized in honour of
 the 70th anniversary of Professor Ji\v{r}\'{i} Niederle at New York University,
 Prague, 5 - 7 May 2009.}} 
\author{Irinel Caprini} 
\affiliation{National Institute of Physics and Nuclear Engineering,  \\Bucharest POB MG-6, R-077125 Romania} 
\author{Jan Fischer}   
\affiliation{Institute of Physics, Academy of Sciences of the Czech Republic,
\\CZ-182 21  Prague 8, Czech Republic}  
\author{Ivo Vrko\v{c}}  
\affiliation{Mathematical Institute, Academy of Sciences of the Czech
Republic, 
\\CZ-115 67  Prague 1, Czech Republic}

\vskip1cm

\begin{abstract}

We address the problem of ambiguity of a function determined by an asymptotic 
perturbation expansion. Using a modified form of the Watson lemma recently 
proved elsewhere, we discuss a large class of functions determined by the same 
asymptotic power expansion and represented by various forms of integrals of 
the Laplace-Borel type along a general contour in the Borel complex plane. 
Some remarks on possible applications in QCD are made.
\end{abstract}

\maketitle

\section{Asymptotic perturbation expansions}
Perturbation expansions are known to be divergent both in quantum
electrodynamics and in quantum chromodynamics, as well as in many other
physically interesting theories and models. In QED, divergence was proved by
F.J. Dyson in 1952 (see \cite{Dyson}). His result has been revisited and
reformulated by many authors (\cite{Lautrup},\cite{tHooft}, see also a review  
in \cite{Jan}). Dyson proposed to give the divergent series mathematical meaning
by interpreting it as an asymptotic series to $F(z)$, the sought function: 
\begin{equation}
F(z) \,\, \sim  \,\, \sum_{n=0}^{\infty} F_{n} z^{n},\quad\quad\quad  
z \in {\cal S}, \quad  z \rightarrow 0,
\label{ptas}
\end{equation} 
where ${\cal S}$ is a point set having the origin as an accumulation point, $z$ 
being the perturbation parameter.  

To see how dramatically the philosophy of perturbation theory changed by this
step, let us first recall the definition of the asymptotic series: 

{\bf Definition:} Let ${\cal S}$ be a region or point set having the origin as 
an accumulation point. The power series $\sum_{n=0}^{\infty}F_{n}z^{n}$ is said 
to be asymptotic to the function $F(z)$ as $z \to 0$ on ${\cal S}$, and we 
write Eq. (\ref{ptas}), if the set of functions $R_{N}(z)$,
\begin{equation}
R_{N}(z) = F(z) - \sum_{n=0}^{N}F_{n}z^{n} ,
\label{rema}
\end{equation}
satisfies the condition
\begin{equation}
R_{N}(z) = o(z^{N}) 
\label{ordo}
\end{equation}
for all $N=0,1,2,...$, $z \rightarrow 0$ and $z \in {\cal S}$.

Note that the asymptotic series is defined by a {\em different limiting 
procedure} than the Taylor one: {\em taking $N$ fixed}, one observes how 
$R_{N}(z)$ behaves for $z \to 0$, $z \in {\cal S}$, the procedure being repeated 
for all $N \geq 0$ integers. Convergence may be provable without knowing $F(z)$, 
but asymptoticity can be tested only if one knows both the $F_{n}$ and $F(z)$. 

By (\ref{ptas}), $F(z)$ is not uniquely determined; there are many different 
functions having the same asymptotic series, (\ref{ptas}) say. The ambiguity of 
a function given by an asymptotic series is illustrated by the lemma of Watson.

\section{Watson lemma}	\label{sec:watlem}
Consider the following integral 
\begin{equation}
\Phi_{0,c}(\lambda) = \int_{0}^{c} e^{-\lambda
x^{\alpha}}\,x^{\beta -1}f(x) {\rm d}x , 
\label{Laplace}
\end{equation}
where $0<c<\infty$ and  $\alpha > 0, \,\beta > 0 $. Let $f(x) \in 
C^{\infty}{[0,c]}$ and $f^{(k)}(0)$  defined as $\lim_{x \to 0+} f^{(k)}(x)$.
Let $\varepsilon$ be any number from the interval $0 < \varepsilon < \pi/2$. 
 
{\bf Lemma 1} (G.N. Watson): {\em If the above conditions are fulfilled,   
the asymptotic expansion}
\begin{equation}
\Phi_{0,c}(\lambda) \sim \frac{1}{\alpha}\sum_{k=0}^{\infty} 
\lambda^{-\frac{k+\beta}{\alpha}}\, \Gamma
\bigg(\frac{k+\beta}{\alpha}\bigg)\frac{f^{(k)}(0)}{k!} 
\label{Watson}
\end{equation}
 {holds \em for $\lambda \rightarrow \infty, 
\lambda \in S_{\varepsilon}$, where  $S_{\varepsilon}$ is the angle}
\begin{equation}
|\arg \lambda| \leq \frac{\pi}{2} - \varepsilon . 
\label{uhel}
\end{equation}
{\em The expansion (\ref{Watson}) can be differentiated with respect to $\lambda$ any number of times.} 

 For the proof see for instance \cite{Jeff}. Let us add several remarks:

{\bf 1)} The angle $S_{\varepsilon}$ of validity of (\ref{Watson}), (\ref{uhel}),
 is independent of $\alpha, \,\beta$  and  $c$.

{\bf 2)} Thanks to the factor $\Gamma \bigg(\frac{k+\beta}{\alpha}\bigg)$,  
the expansion coefficients in (\ref{Watson}) grow faster with $k$ than those 
of the Taylor series for $f(x)$. 

{\bf 3)}  The expansion coefficients in (\ref{Watson}) are  independent of $c$. 
This illustrates the impossibility of a unique determination of a function from 
its asymptotic expansion.	

In the next section we shall give a modification to the Watson lemma, which 
shows that under plausible assumptions the straight integration contour can be 
bent. 	

\section{Modified Watson lemma}	\label{sec:modwat}
The modified Watson lemma we present below (and call Lemma 2') is a special 
case of Lemma 2, which we publish and prove in Ref. \cite{9watson}. The special 
form given here is obtained from that given in \cite{9watson} by setting 
$\alpha=\beta=1$. 
  
Let  $G(r)$  be a continuous complex function of the form $G(r) = 
r \exp (ig(r))$, where $g(r)$ is a real-valued function given on $0\le r < c$, 
with   $0<c \leq \infty$. Assume that the derivative $G'(r)$ is continuous on 
the interval $0\le r < c$ and a constant $r_0>0$ exists such that 
\begin{equation}
|G'(r)| \le K_1 r^{\gamma_1}, \quad\quad r_0\le r < c,
\label{CK1}
\end{equation}
for a nonnegative $K_1$ and a real $\gamma_1$.

Assume that the parameter $\varepsilon >0$ exists such that the quantities 
\begin{equation}
A = \inf_{r_0\leq r < c} g(r),\quad\quad \quad B = 
\sup_{r_0 \leq r < c} g(r)
\label{infsup} 
\end{equation}
satisfy  the inequality
\begin{equation}
B-A < \pi-2\varepsilon.
\label{eps} 
\end{equation}

Let the function $f(u)$ be defined along the curve $u=G(r)$ and on the disc 
$|u|<\rho$, where $\rho>r_0$. Let $f(u)$ be holomorphic on the disc 
and measurable on the curve. Assume that  
\be
|f(G(r))|\leq K_2 r^{\gamma_2},\quad\quad r_0\le r < c,
\label{CK2}\end{equation}
hold for a nonnegative $K_2$ and a real $\gamma_2$.

Define the function  $\Phi_{b,c}^{(G)}(\lambda)$ for $0 \leq b< c$
by\footnote{This integral exists since we assume that $f(u)$ is measurable along
the curve $u=G(r)$ and bounded by (\ref{CK2}).}  
\begin{equation}
\Phi_{b,c}^{(G)}(\lambda)= \int_{r=b}^{c} e^{-\lambda G(r)} 
G(r) f(G(r)) dG(r).
\label{bc}
\end{equation}

{\bf Lemma 2':} {\em If the above assumptions are fulfilled, then 
the asymptotic expansion}
\begin{equation}
\Phi_{0,c}^{(G)}(\lambda) \sim  \sum_{k=0}^\infty \lambda^{-(k+1)}\, 
\Gamma(k+1)
\frac{f^{(k)}(0)}{k!}  
\label{V}
\end{equation}
 {\em holds for $\lambda \rightarrow \infty, \lambda \in \cal T_{\varepsilon}$, where}
\begin{equation}
{\cal T}_\varepsilon = \{\lambda: \lambda=|\lambda| \exp({\rm i} \varphi), \, \, \,  
- \frac{\pi}{2}- A + \varepsilon <\varphi< \frac{\pi}{2} - B- \varepsilon  \}.
\label{calT} 
\end{equation}

We refer the reader to Ref. \cite{9watson} for the proof of Lemma 2 and its 
discussion. The above simplified version, Lemma 2', is given here to illustrate 
some special features of the general Lemma 2 and its possible applications. 

Let us add several remarks to Lemma 2':

{\bf 1/} Lemma 2' implies Watson's lemma when the integration contour is chosen
to have the special form of a segment of the real positive semiaxis, 
{\em i.e.} $g(r)\equiv 0$, and $f(r) \in C^\infty[0,c]$. 

{\bf 2/} Perturbation theory is obtained by setting $\lambda=1/z$  
in  (\ref{CK2}), (\ref{bc}). Then, the function
\begin{equation}
F^{(G)}_{0,c}(z)= 
\int_{r=0}^c e^{-G(r)/z}\, f(G(r))\, dG(r) 
\label{abz11}
\end{equation}
has the asymptotic expansion
\begin{equation}
F_{0,c}^{(G)}(z) \sim \sum_{k=0}^\infty z^{k+1} f^{(k)}(0)  
\label{asyF} 
\end{equation}
 for $z \rightarrow 0$ and $z \in \cal Z_\varepsilon$, where 
\begin{equation}
{\cal Z}_\varepsilon =\{z: z=|z| \exp{(i \chi)},\, -\frac{\pi}{2}+B + \varepsilon <\chi< 
\frac{\pi}{2}+A - \varepsilon \}.
\label{calZ}
\end{equation}

{\bf 3/} The parameter $\varepsilon$ in (\ref{eps}) is limited by 
$0< \varepsilon<\pi/2-(B-A)/2$, but is otherwise arbitrary. Note however that 
the upper limit of $\varepsilon$ depends on $B-A$ and may be considerably less 
than $\pi/2$. This happens, for instance, if the integration contour is 
bent or meandering.

{\bf 4/}  The parametrization $G(r)=r \exp{(ig(r))}$ does not include contours 
that cross a circle centred at $r=0$, either touching or doubly intersecting 
it, so that the derivative $G'(r)$ either does not exist or is not bounded. 
In such cases, the parametrization has to be modified.  

{\bf 5/} 
Let us remark that the proof of Lemma 2 in Ref. \cite{9watson} allows us to 
obtain remarkable correlations between the strength of the bounds on the 
remainder and the size of the angles within that the asymptotic expansion is 
valid. It follows from \cite{9watson} that the bounds are proportional to 
\begin{equation}
\frac{1}{(|\lambda|-1) \sin \varepsilon} {\rm e}^{-(|\lambda|-1)
r_0 \sin \varepsilon}
\label{bound1}
\end{equation}
or to
\begin{equation}
C_{N} (|\lambda| \sin\varepsilon)^{-(N+ 2)}  ,
\label{bound2}
\end{equation}
where $N$ is the truncation order and the $C_{N}$, $N=0, 1, 2, ...$ are 
$\lambda$-independent positive numbers. The bounds decrease with increasing 
$\varepsilon$, the parameter, which  determines the angles  
$\cal T_\varepsilon$ and $\cal Z_\varepsilon$, see (\ref{calT}) and 
(\ref{calZ}) respectively. As a consequence, the larger the angle of validity, 
the looser the bound, and vice versa.  
 \section{Some applications to perturbative QCD} \label{sec:remqcd}
To discuss some applications of Lemma 2', we take the  Adler function 
\cite{Adler},    
\begin{equation}
{\cal D}(s) =- s\frac{{\rm d}\Pi(s)}{{\rm d}s}-1\,.
\label{calD}
\end{equation}
where $\Pi(s)$ is the polarization amplitude defined in terms of the vector 
current products for light quarks. The Adler function ${\cal D}(s)$ is real 
analytic in the $s$-plane, except a cut along the timelike axis produced by 
unitarity \cite{BogSh, Adler}. In perturbative QCD, any finite-order 
aproximant has cuts along the timelike axis, while the 
renormalization-group improved expansion,
\be\label{Dpert}
{\cal D}(s) = D_1 \,\alpha_s(s)/\pi +  D_2 \,(\alpha_s(s)/\pi)^2 + 
D_3 \,(\alpha_s(s)/\pi)^3  + \ldots \,,  
\ee 
has, in addition,  an unphysical singularity due to the Landau pole in the 
running coupling $\alpha_s(s)$. (\ref{Dpert}) is known to be divergent, the 
$D_n$ growing as $n!$ at large orders \cite{Mueller1992}-\cite{BenekePR}. 

\subsection{On the high ambiguity of perturbative QCD}
To discuss the implications of Lemma 2', we first define the  Borel transform 
$B(u)$  by  \cite{Neubert1996}, 
\be\label{B}
B(u)= \sum\limits_{n\ge 0} b_n \,u^n,\quad\quad \quad 
b_n=\frac{D_{n+1}}{\beta_0^n\,n!}\,.
\ee
It is usually assumed that the series (\ref{B}) is convergent on a disc of
nonvanishing radius (this result was rigorously proved by David et al. 
\cite{David} for the scalar $\varphi^4$ theory in four dimensions). This is 
what is required in Lemma 2' for the generalized Borel transform $f(G(r))$.

If we assume that the series (\ref{Dpert}) is asymptotic, Lemma 2'  
implies a large freedom in recovering the true function from  
its  coefficients. 
All the  functions ${\cal D}^G_{0,c}(s)$ of the form
\be\label{BDG}
{\cal D}^G_{0,c}(s)=\frac{1}{\beta_0} \int_{r=0}^c e^{-\frac{G(r)}{\beta_0\,
a(s)}}\, B(G(r))\, {\rm d}G(r) \, ,
\ee  
where $a(s)=\alpha_s(s)/\pi$,  admit the asymptotic expansion
\be\label{DGpert}
{\cal D}^G_{0,c}(s) \sim \sum\limits_{n=1}^{\infty} D_n \, (a(s))^n, \quad 
\quad \quad  a_{s}(s) \to 0, 
\ee
in a certain domain of the $s$-plane, which follows from (\ref{calT}) and the 
expression of the running coupling $a(s)$ given by the renormalization group.
No function of the form ${\cal D}^G_{0,c}(s)$, (\ref{BDG}), can be a  priori 
preferred when looking for the true  Adler function. 

Contributing only to the exponentially suppressed remainder, neither the 
form or length of the contour, nor the values of $B(u)$ outside the 
convergence disc can affect (\ref{DGpert}).  
The remainder to (\ref{DGpert}) is of the form  $h\, {\rm 
exp}(- d/\beta_{0} a(s)) \sim h \left(-\Lambda^{2}/s \right)^{d}$. The 
quantities $h$ and $d>0$  depend on the contour and on $B(u)$ outside the disc, 
which can be chosen rather freely. As a consequence, (\ref{BDG}) contains 
arbitrary power terms, to be added to (\ref{DGpert}).  

\subsection{Analyticity and optimal conformal mapping}

In discussing the divergence of (\ref{Dpert}) and (\ref{B}), the singularities 
of ${\cal D}(s)$ in the $\alpha_s(s)$ plane and, respectively, those of $B(u)$ 
in the Borel plane are of importance. As for $B(u)$, some information about the 
location and nature of the singularities can be obtained from certain classes 
of Feynman diagrams (which can be summed, see 
\cite{Beneke1993}-\cite{BenekePR}), and from general arguments based on 
renormalization theory, \cite{Mueller1992,BeBrKi}. It follows that $B(u)$  
has branch points along the rays $u \geq 2$ and $u\leq -1$ (IR and UV 
renormalons respectively). Other (though nonperturbative) singularities, 
for $u \ge 4$, are produced by instanton-antiinstanton pairs. (Due to 
the singularities at $u>0$, the series (\ref{Dpert}) is not Borel summable.) 
No other singularities of $B(u)$ in the Borel plane are known, however. It 
is usually assumed that $B(u)$ is holomorphic elsewhere. 

To make full use of analyticity of $B(u)$ in the whole ${\cal B}$, we shall use 
the method of optimal conformal mapping \cite{SorJa}. Let ${\cal K}$ be the 
disc of convergence of the series (\ref{B}); clearly, 
${\cal K} \subset {\cal B}$. Then, evidently, the expansion (\ref{B}) in powers of $u$ can be 
replaced by that in powers of $w(u)$,
\be\label{Bw}
B(u)=\sum_{n\ge 0} c_n \,w^n,
\ee
where the function $w=w(u)$ with the property $w(0)=0$ represents the conformal
mapping of the region of ${\cal B}$ onto the disc $|w|<1$, on which (\ref{Bw}) 
converges. It can easily be seen that (\ref{Bw}) has better convergence 
properties than (\ref{B}) in this case: indeed, as was proved in \cite{SorJa} by
using Schwarz lemma, the larger the region mapped by $w(u)$ onto $|w|<1$, the 
faster the large-order convergence rate of (\ref{Bw}). 

If $w(u)$ maps the whole ${\cal B}$ onto the unit disc $|w|<1$ in the $w$ plane, 
the mapping is called optimal. In this case, (\ref{Bw}) converges everywhere on 
${\cal B}$ and the convergence rate is the fastest \cite{SorJa}.  	
The region of convergence of (\ref{Bw}) coincides with ${\cal B}$, the region 
of analyticity. In this way, the optimal conformal mapping can express 
analyticity in terms of convergence. 

Inserting (\ref{Bw}) into (\ref{BDG}) we obtain an alternative asymptotic
expansion: 
\be\label{BDGw}
{\cal D}^G_{0,c}(s)=\frac{1}{\beta_0} \int_{r=0}^c e^{-\frac{G(r)}{\beta_0\,
a(s)}}\, \sum_{n\ge 0} c_n \, [w(G(r))]^n \, {\rm d}G(r) \,.
\ee  
Containing powers of the optimal conformal mapping $w(u)$ (which has the same location 
of singularities as the expanded function $B(u)$), this representation implements more 
information about the singularities of $B(u)$ than the series (\ref{B}) in powers of 
$u$, even at finite orders. Thus, it is to be expected that even the finite-order 
approximants of (\ref{BDGw}) will provide a more precise description of the  function 
searched for \cite{CaFi, CvLe}. 

\subsection{Analyticity may easily get lost}
We shall shortly mention an intriguing situation showing that a careless
manipulation with the integration contour may have a fateful impact on 
analyticity. In \cite{HoMa}, two different integration contours in the 
$u$-plane were chosen for the summation of the so-called renormalon chains 
\cite{Beneke1993}: for $a(s)>0$ and $a(s)<0$, a ray parallel and close to the 
positive and, respectively, negative semiaxis is chosen.   
As was expected and later proved \cite{CaFi4}, analyticity is lost with this 
choice, the summation being only piecewise analytic in $s$. 

On the other hand, as shown in \cite{CaNe, CaFi3}, the Borel summation with the 
Principal Value (PV) prescription of the same class of diagrams  admits an 
analytic continuation in the $s$-plane, in consistence with analyticity except 
a cut along a segment of the spacelike axis, related to the Landau pole.

\section{In conclusion} \label{sec:conrem} 
In this talk we discussed some special consequences of our general result 
published in \cite{9watson}, which is based on a modification of Watson 
lemma. It follows that a perturbation series, if regarded as asymptotic, 
implies a huge ambiguity of possible expanded functions having the same 
asymptotic expansion of the type (\ref{ptas}). This mathematical fact is 
often ignored or overlooked in physical applications. Our contribution consists 
in the fact that we have specified its special subclass by Lemma 2 of Ref. 
\cite{9watson}. Moreover, in the present talk, we consider a special subclass 
of Lemma 2 (as defined by Lemma 2' in section III of this talk), which we
discuss here in more detail due to its direct applicability to perturbative QCD.  
To find the true solution, additional information inputs are unavoidable. 

Applying the result to QCD, we conclude that the contour of the integral 
representing the QCD correlator can be chosen very freely. The same holds for 
the Borel transform $B(u)$  outside the convergence circle. 

We kept our discussion on a general level, bearing in mind that little is known, 
in a rigorous framework, about the analytic properties of the QCD correlators in 
the Borel plane. If some specific properties are known or assumed, the integral 
representations will have additional analytic properties. Naturally, the results
obtained in \cite{9watson} may also be useful in other branches of physics 
where perturbation series are divergent. 

\begin{acknowledgments} 
One of us (I.C.) thanks Prof. J. Ch\'yla and the Institute of Physics of the 
Czech Academy in Prague for hospitality. J.F. thanks Prof. P. R\c aczka and the 
Institute of Theoretical Physics of the Warsaw University for hospitality. 
Supported by CNCSIS in the frame of the Program Idei, Contract Nr. 464/2009, 
and by the Projects No. LA08015 of the Ministry of Education and 
AV0-Z10100502 of the Academy of Sciences of the Czech Republic.
\end{acknowledgments}


\end{document}